\documentclass[trackchanges]{aastex701}

\usepackage{graphicx,times}             
\usepackage{color}
\usepackage{ulem}
\usepackage{multirow}
\usepackage{rotating}
\usepackage{CJK}

\newcommand{\shao}{Shanghai Astronomical Observatory, Chinese Academy of Sciences, Shanghai 200030, China}
\newcommand{\CAS}{Chinese Academy of Sciences, University of Chinese Academy of Sciences, 19A Yuquan Road, Beijing 100049, China}

\begin{document}
\begin{CJK*}{UTF8}{gbsn}
\title{A 0.03 Hz Radio Quasi-periodic Oscillation During the 2025 Flare of GRS 1915+105}

\author{Ya Xing Li (李亚星)}
\affiliation{\shao}
\affiliation{\CAS}
\email{yxli@shao.ac.cn}

\author[0000-0002-2920-1880]{Lei Liu (刘磊)}
\affiliation{\shao }
\email[show]{liulei@shao.ac.cn}

\author[0000-0001-7369-3539]{Wu Jiang (江悟)}
\affiliation{\shao}
\email[show]{jiangwu@shao.ac.cn}

\author[0000-0002-5385-9586]{Zhen Yan (闫震)}
\affiliation{\shao }
\email[show]{zyan@shao.ac.cn}

\author{Bo Xia (夏博)}
\affiliation{\shao }
\email{bxia@shao.ac.cn}

\author{Zhi Qiang Shen (沈志强)}
\affiliation{\shao }
\email{zshen@shao.ac.cn}

\correspondingauthor{Lei Liu, Wu Jiang, Zhen Yan}

\begin{abstract}

Our weekly-cadence radio monitoring campaign captured a bright flare in 2025 from the microquasar GRS 1915+105, observed simultaneously in the S- and X-bands (2.25 GHz and 8.42 GHz) with a short single baseline of two radio telescopes in Shanghai. Through high time resolution analysis, we detected a significant and short-lived quasi-periodic oscillation (QPO) at $\sim$0.03 Hz and its harmonic ($\sim$0.06 Hz) in both radio bands of two consecutive observations on MJD 60765 ($>5.9 \sigma$) and MJD 60772 (2.8$\sigma$). Crucially, the QPO frequency is identical in both radio bands and matches oscillations detected in previous years. The recurrence and wavelength independence of the QPO frequency suggest an intrinsic characteristic timescale of the accretion-jet system.

\end{abstract}

\keywords{  \uat{Black holes} {162},\uat{Radio jets}{1347},\uat{X-ray binary stars}{1811}}


\section{Introduction}

GRS 1915+105, a prominent Galactic microquasar, was first detected by the WATCH all-sky monitor on 1992 August 15 \citep{Castro-Tirado1992}. It became the first known Milky Way source to exhibit superluminal radio jets \citep{MR1994Na}. For more than two decades, GRS 1915+105 remained in a persistently high-luminosity state, unlike the episodic outbursts typical of classical transient black hole X-ray binaries (BHXRBs). Its X-ray spectral and timing behavior can be described by three states, where State C resembles the canonical hard state in BHXRBs, while State A and B are softer \citep{Belloni2000}. Both GRS 1915+105 and classical transient BHXRB exhibited analogous jet activities: a hard X-ray state linked to a steady, compact radio jet, and transient, optically thin ejections during state transitions\citep{FB2004ARAA,Belloni2010}.


A significant turning point occurred after 2019 when GRS 1915+105 entered an unprecedented obscuration phase phase characterized by an X-ray flux drop of one to two orders of magnitude \citep[]{Neilsen018,Neilsen2020,Miller2020}. This phase is interpreted as the dense, large-scale absorber that shrouds the inner accretion flow, effectively blocking X-ray emission from the central engine \citep{Neilsen2020,Miller2020,Gandhi2025MNRAS,Miller2025}.
Despite this, the source remained radio-luminous and highly variable, displaying multiple flares from 2019 to 2021, followed by a two-year low-flux period \citep{Motta2021,Sanchez-Sierras2023}. Two even brighter radio flares occurred in 2023 \citep{Trushkin_2023ATel15964,Trushkin_2023ATel15974,Trushkin_2023ATel16168}, accompanied by near-simultaneous infrared brightening \citep{Sanchez-Sierras2023,Gandhi2025MNRAS}. The persistence of radio and infrared activity during this phase suggests that the central engine remained active but heavily obscured \citep{Motta2021,Sanchez-Sierras2023,Gandhi2025MNRAS}, challenging conventional views of disk wind and jet \citep{Neilsen2009Nat}.   

Since its discovery, GRS 1915+105 has displayed extreme and rich variability across X-ray \citep[e.g. ][]{Trudolyubov1999,Lee2002,Negoro2018,Miller2020,Zhou2025}, infrared \citep[e.g. ][]{Eikenberry2000,Fuchs2003}, and radio emissions \citep[e.g. ][]{PF1997MNRAS,Fender1999,Vadawale2003,RM2025ApJ}. Understanding jet behavior requires analyzing radio variability over a wide range of timescales. Historically, GRS 1915+105 has exhibited radio quasi-periodic oscillations (QPOs) on 20--40 min timescales \citep{PF1997MNRAS}, interpreted as evidence linking disk instabilities to jet ejections \citep{Mirabel1998}. Using FAST, \citet{Tian2023} recently achieved the first detection of a subsecond ($\sim$5 Hz) radio QPO in GRS 1915+105, suggesting rapid oscillation within the relativistic jet. Such findings significantly expand our understanding of jet variability and highlight the importance of probing the innermost jet launching region through radio timing on short-timescale.

To investigate the jet activity of GRS 1915+105 in its current X-ray obscuration phase, we conducted a dual-band radio monitoring campaign routinely with the short single baseline of two radio telescopes in Shanghai. Here we report the observations of the bright radio flare in 2025. The subsequent sections are organized as follows: Section 2 outlines the radio and X-ray observations and data reduction procedures; Section 3 presents our primary findings; Section 4 offers a discussion on the physical properties of the radio flare and the potential origin of the identified $\sim$0.03 Hz radio oscillation; and Section 5 provides a summary of our findings.

\section{Observation and Data Analysis}

\subsection{ARM Observation}

The Automatic Radio-transient Monitoring (ARM) project, supported by the Shanghai Astronomical Observatory (SHAO), is designed to conduct routine radio observations of flares and outbursts from astronomical transient sources. The technique of Very Long Baseline Interferometry (VLBI) is utilized by using the approximately $6\text{ km}$ baseline between the Tianma $65\text{m}$ and Sheshan $25\text{m}$ radio telescopes located in Shanghai. The project systematically monitors flux variability on a weekly cadence in two bands simultaneously: S-band (2.25 GHz, 96 MHz bandwidth) and X-band (8.42 GHz, 320 MHz bandwidth).

ARM has monitored GRS 1915+105 since November 2024, typically observing the source weekly. Each observation session consists of 1 to 5 target scans. To ensure accurate calibration, these target scans are preceded and/or followed by $\sim$2 min scans of a calibrator (B1919+086 or TXS 2013+370). The target exposure per scan ranged from 12--23 min, except for three extended 111 min scans near the flare peak. On MJD 60737, the S-band bandwidth was changed to 64 MHz. During 172 days, a total of 61 scans from 24 observations of GRS 1915+105 were analyzed.

\subsection{Radio flux reduction}
Given the complexity of VLBI data processing, a dedicated pipeline based on DiFX \citep{Deller2007PASP} was developed within the ARM project to perform correlation, fringe fitting, and data archiving/filtering. Each observation is processed in two runs: the first includes only calibrator scans to extract the X-band Single Band Delay (SBD) for clock correction, and the second processes all scans using the calibrated clock offset and rate.

The flux density of each source is derived from the baseline signal-to-noise ratio (SNR), obtained directly from the \texttt{fourfit} output, along with the system equivalent flux density (SEFD), antenna gains ($g$), bandwidth ($\Delta\nu$), and integration time ($t_\mathrm{int}$):
\begin{equation}
    F=\frac{\mathrm{SNR}}{\sqrt{g_1\cdot g_2}}\cdot\sqrt{\frac{\mathrm{SEFD}_1\cdot \mathrm{SEFD}_2}{2\Delta\nu~t_\mathrm{int}}}.\label{eq:flux}
\end{equation}
The antenna gain is derived from elevation-dependent gain curves. The SEFD is calculated as ${2kT_\mathrm{sys}}/{\eta A}$, where $A$ is the antenna area, $\eta$ is the efficiency, and $T_\mathrm{sys}$ is the mean system temperature over all frequency channels. For the Sheshan 25 m telescope, S-band 
$T_\mathrm{sys}$ necessitated adopting a fixed default value of 85.4 K due to possible severe RFI contamination.

Fluxes from Eq.~\ref{eq:flux} are further calibrated against amplitude calibrators to mitigate instrumental fluctuations. If $F_{\mathrm{ref,obs}}$ and $F_{\mathrm{obs}}$ are the observed fluxes of the calibrator and target, and $F_{\mathrm{ref,std}}$ is the catalogued calibrator flux from the Radio Fundamental Catalogue (RFC)\footnote{\url{http://astrogeo.org/}}, the calibrated target flux is
\begin{equation}
    F_{\mathrm{cal}} = F_{\mathrm{obs}} \frac{F_{\mathrm{ref,std}}}{F_{\mathrm{ref,obs}}}.
\end{equation}
The corresponding flux uncertainty combines measurement and calibration errors:
\begin{equation}
    \sigma_F = \frac{|F_{\mathrm{ref,std}} - F_{\mathrm{ref,obs}}|}{F_{\mathrm{ref,std}}} F_{\mathrm{cal}}.
\end{equation}

Initially, TXS~2013$+$370 was used as the calibrator, but after MJD~60691 it was replaced by B1919$+$086. In the observation on MJD~60646, where TXS~2013$+$370 yielded no detectable fringes, the nearby galaxy M81 was adopted as a suitable substitute. The extended 111\,min scan on MJD~60691 was divided into three $\sim$2,200\,s segments to facilitate processing.

\subsection{NICER data reduction and analysis}

We reproduced cleaned events with the \texttt{nicerl2} task from the NICER-specific tools of HEASOFT v6.33 using the calibration dataset \texttt{xti20240206}. Good Time Intervals (GTIs) were selected with the following criteria: elevation above the Earth limb $>30^{\circ}$, angular distance from the bright Earth $> 40^{\circ}$, overshoot rate $<$ 10 cts per module, undershoot rate $<$ 200 cts per module, and geomagnetic cutoff rigidity $>2$ GeV. Default noisy detectors were excluded. Spectra and light curves were extracted within these GTIs using \texttt{nicerl3-spect} and \texttt{nicerl3-lc} with the SCORPEON background model. The spectra were regrouped with at least 25 counts per bin.

Spectral fitting was performed in the 1.5--10~keV band for data between MJD~60720 and MJD~60820 using \texttt{XSPEC} v12.12.1 with the \texttt{chi} statistic. The adopted model was \texttt{TBabs*TBpcf*powerlaw}. Given that GRS~1915$+$105 is in an obscuration phase \citep{Neilsen2020,Miller2020}, we included both interstellar absorption (\texttt{TBabs}) and partial covering absorption (\texttt{TBpcf}). The interstellar hydrogen column density was fixed at $N_\mathrm{H}=5\times10^{22}\ \mathrm{cm^{-2}}$ \citep{Lee2002}. Two or three Gaussian components were added to model the broad and narrow Fe emission lines \citep{Neilsen2020,Miller2020,Zhou2025}. The 1--10~keV flux was obtained using the \texttt{cflux} convolution model.

\section{Result}
\subsection{The long-term evolution of 2025 radio flare}

\begin{figure*}
\centering
\includegraphics[width=0.9\linewidth]{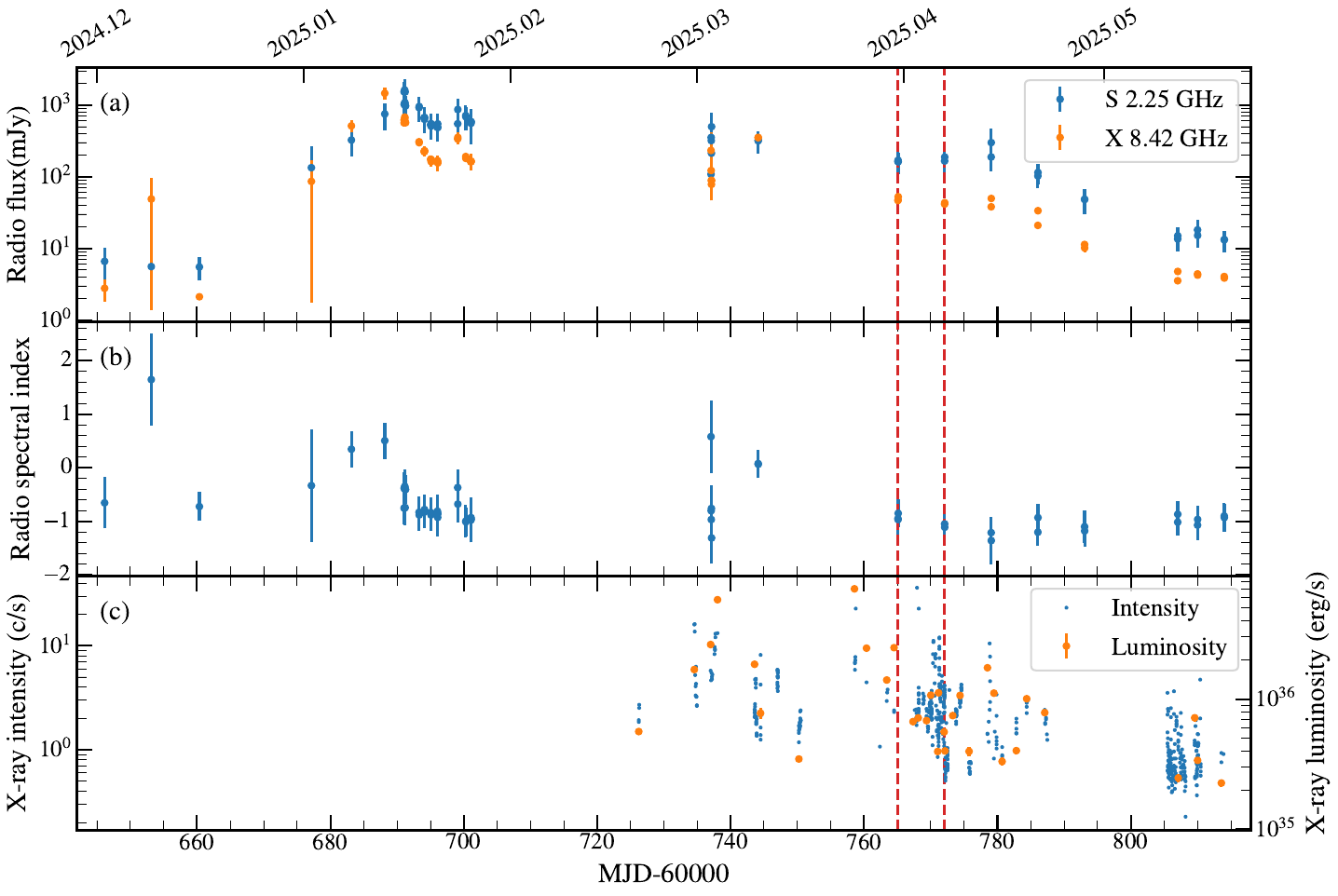}
\caption{The radio and X-ray lightcurves of the 2025 flare from MJD 60646 to MJD 60814. The two red dashed lines mark the observation dates with QPO detection.  (a) The radio flux densities at 2.25 and 8.42 GHz. (b) The radio spectral index. (c) The blue points represent the $NICER$ light curve of count rate and the orange points represent the luminosity in energy band 1--10 keV.
\label{ref:flare}}
\end{figure*}

The flux evolution in both radio bands during the 2025 flare is shown in Figure~\ref{ref:flare}a, and the corresponding spectral index $\alpha$, derived from $S_{\nu} \propto \nu^{\alpha}$, is shown in Figure~\ref{ref:flare}b. The flux began to rise after MJD~60660, reaching a sharp peak shortly thereafter. The X-band flux peaked at $\sim$1461~$\pm$~283~mJy on MJD~60688, while the S-band reached its maximum of 1624~$\pm$~667~mJy about three days later, on MJD~60691 \citep[see also][]{Trushkin2025}. Around MJD~60690, the spectral index decreased from $+0.5$ to $-0.75$, indicating a transition from an optically thick to optically thin jet spectrum. After the peak, the fluxes in both radio bands declined gradually, accompanied by substantial variability in the spectral index during these episodes.

The $NICER$/XTI light curve covering MJD~60646--60814 (Figure~\ref{ref:flare}c) corresponds to the period of gradual radio decay. The X-ray luminosity during this period was in the range $1\times10^{35}$ -- $7\times10^{36}\ \mathrm{erg\ s^{-1}}$. Due to Sun constraints, no NICER observations were available during the rise or early decay of the radio flare. The X-ray count rate remained low relative to the pre-2019 level, exhibiting strong variability on timescales of days or less, superposed on a long-term decline. Given the limited cadence of our radio monitoring, short-timescale ($\lesssim$ 7 days) radio flares were likely missed, suggesting that the radio emission may also vary significantly on similar timescales.

\subsection{Short term variability of radio emission}

\begin{figure*}
\includegraphics[width=\linewidth]{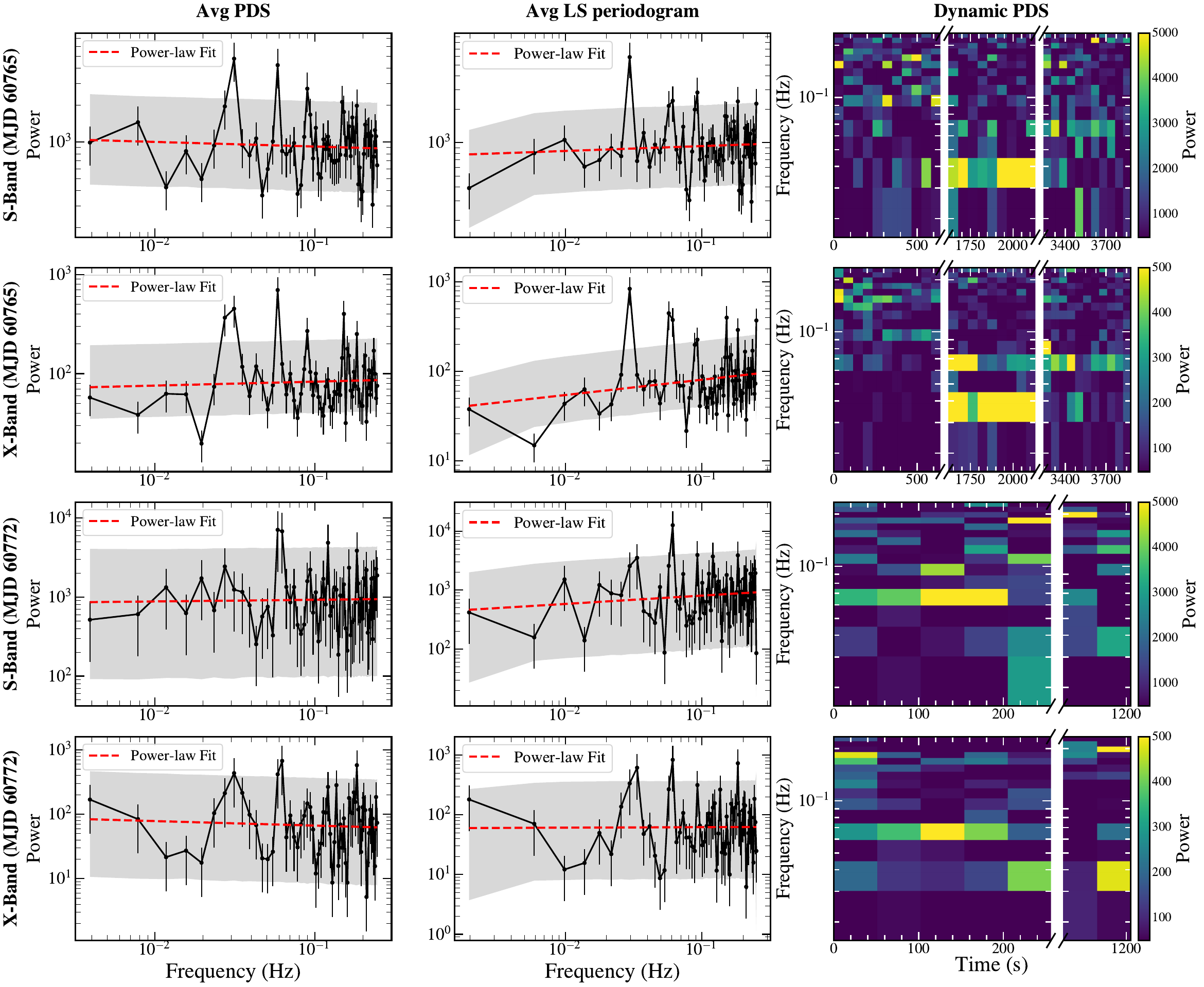}
\caption{Averaged PDS, Averaged LS periodogram and dynamic PDS of the observations on MJD 60765 (top two rows) and MJD 60772 (bottom two rows). The first and third rows present the results for the S-band, while the second and fourth rows present the X-band results. For each observation date and band, the left column displays the observed averaged PDS (black data points with error bars), the middle column displays the averaged LS periodogram (black data points), and the right column shows the corresponding dynamic PDS. In the averaged PDS and LS periodogram panels, the red dashed line represents the best-fit power-law model, and the grey shaded region represents the $1\%-99\%$ percentile range of the simulated power at each frequency. This percentile demonstrates the typical power level reached by most simulated PDSs and provides a visual comparison with the observed PDS. It is not used to define the significance of the QPO detection. In addition, the apparent peaks observed above 0.06 Hz are consistent with being harmonics of the fundamental 0.03 Hz. The white vertical blank regions in the dynamic PDS correspond to the time intervals between individual observation scans used for flux calibration.}
\label{ref:pdsallnew}
\end{figure*}

\begin{figure*}
\includegraphics[width=\linewidth]{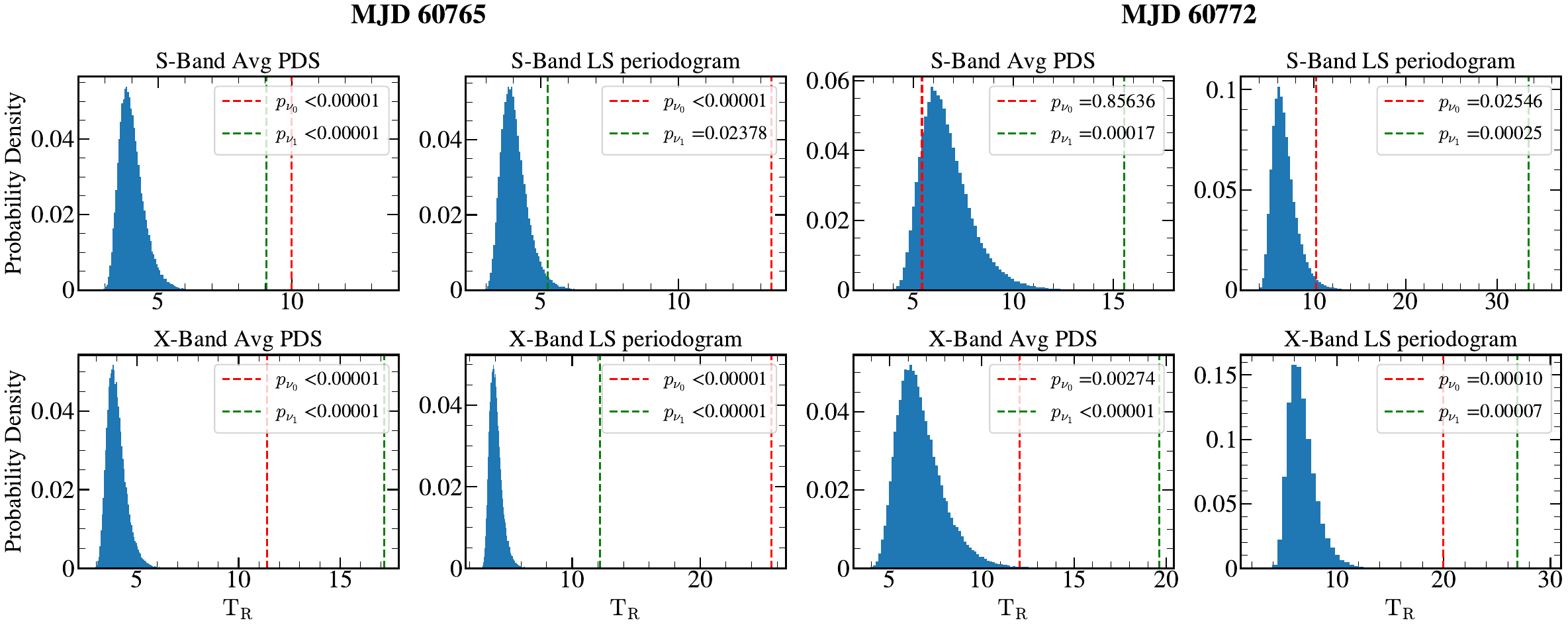}
\caption{The distributions of the maximum $T_{\mathrm{R}}$ from each simulated data for the corresponding wavebands. $T_{\mathrm{R}}=\mathrm{max}(2I_j/S_j)$ is calculated from each simulated PDS and LS periodogram, where $I_j$ is the power at frequency $f_j$ and $S_j$ is the best-fit power-law value at that frequency. The red and green dashed lines indicate $T_{\mathrm{R}}^{\mathrm{obs}}$ at the fundamental and harmonic frequencies, respectively.} \label{ref:hist}
\end{figure*}

Our data processing pipeline allows flux measurements on timescales of seconds by adjusting the fitting window boundaries for a specific temporal resolution. We then generated 2\,s resolution light curves for each scan. All time bins within a single scan share a common calibration uncertainty determined from the entire integral time of the scan.

The corresponding power density spectrum (PDS) of each scan is calculated using the \texttt{Stingray} package \citep{Huppenkothen2019ApJ}. Most PDSs follow a power-law shape with very small index, indicating red or white noise dominated variability. However, in several scans obtained on MJD~60765 and MJD~60772, a prominent peak appeared near 0.03~Hz in both S- and X-band PDSs, accompanied by a harmonic at 0.06~Hz. No other significant peaks were found in the PDSs of the remaining epochs.

Dynamic PDSs trace the temporal evolution of the QPO. The 0.03~Hz feature and its harmonic become prominent during the second scan on MJD~60765, with the harmonic dominating in the third scan on MJD~60765 and in both scans on MJD~60772 (see \autoref{ref:pdsallnew}), suggesting a transient nature for the 0.03 Hz signal and its harmonic.

To further analyze these QPO signals, we combined the scans for both observation dates to produce averaged PDSs with 256~s segments, and fitted them with a model consisting of a power-law continuum plus Lorentzian components.
The combination of these models is commonly applied for PDS fitting in X-ray timing analysis \citep{Belloni2002ApJ}, where the power-law component describes broad-band noise and the Lorentzian component describes the QPO. The best-fit power-law indices are close to zero, and significant Lorentzians are required at the fundamental ($\nu_0\sim0.03$~Hz) and second harmonic ($\nu_1\sim0.06$~Hz) frequencies, except for the S-band PDS on MJD~60772, where only the harmonic is evident (see \autoref{ref:pdsallnew}). The detected QPOs are narrow, with quality factors $Q=\nu_0/\mathrm{FWHM}>5$, while the harmonics are even sharper ($Q>20$). 
Radio QPOs at the same frequency were previously reported by \citet{Wang2025NC} in FAST radio data from January~2021, suggesting a recurrent phenomenon.

To assess the statistical significance of these detections, we followed the Bayesian approach of \citet{Vaughan2010MNRAS}. We simulated 100,000 light curves per radio band using the \texttt{colorednoise}\footnote{\url{https://github.com/felixpatzelt/colorednoise}} code \citep{Timmer1995}, adopting the best-fit power-law indices of the averaged PDSs, which could reproduce the stochastic fluctuations of power-law noise. Crucially, to account for the procedure that all the observations in this flare were searched for QPO, each simulated light curve was generated with a duration of 76,000 s, approximately matching the total 2025 observing time. From each simulated light curve, we randomly extracted a segment with the same length as the observation with QPO detected. We then computed the PDS from these extracted segments, refitted each PDS with a power-law model and computed the test statistic $T_{\mathrm{R}}=\mathrm{max}(2I_j/S_j)$ for each simulation, where $I_j$ is the power at frequency $f_j$ and $S_j$ is the best-fit power-law value at that frequency. 
Here, the ratio $2I_j/S_j$ quantifies the power excess relative to the underlying power-law continuum at a given frequency. The value $T_{\mathrm{R}}$ derived from each simulation therefore represents the maximum fluctuation that can be produced by power-law noise over the full searched frequency range. We then compared the observed $T_{\mathrm{R}}^{\mathrm{obs}}$ at 0.03~Hz and 0.06~Hz with the posterior distribution of $T_{\mathrm{R}}$ to obtain the $p$-value (Figures~\ref{ref:hist}). The $p$-value is defined as the fraction of simulations with $T_{\mathrm{R}} > T_{\mathrm{R}}^{\mathrm{obs}}$, and represents the false-alarm probability that power-law noise alone could produce a peak at least as strong as the observed signal. Finally, the $p$-value was converted into an equivalent Gaussian sigma significance for consistency with the conventional notation.
Since the S- and X-band observations are simultaneous and independent, we combined their probabilities using Fisher's method. The results are summarized in Table~\ref{tab:QPOsigma}. On MJD~60765, the 0.03~Hz signal is detected with $>4.3\sigma$ significance in both S- and X-bands and the corresponding combined significance is $>5.9\sigma$. On MJD~60772, the fundamental QPO is marginally significant ($2.8\sigma$) only in the X-band, while the harmonic at 0.06 Hz is robustly detected in both bands ($3.6\sigma$ in X-band and $>4.3\sigma$ in S-band), reaching a joint significance of $>5.4\sigma$. 
In addition, several smaller but prominent peaks visible in the averaged PDSs align with the positions of higher-order harmonics (see \autoref{ref:pdsallnew}).

Furthermore, we also performed Lomb-Scargle (LS) periodogram analysis to verify the reliability of the QPO signals. The LS periodogram is a statistical tool specifically designed to detect periodic signals in unevenly sampled light curves \citep{VanderPlas2018}. Using the same 256 s segmentation and similar Bayesian method as described above, we generated the averaged LS periodograms and evaluated the signal significance. As shown in Figures~\ref{ref:pdsallnew} and~\ref{ref:hist}, the QPOs remain highly significant, confirming the robustness of our detections.

\begin{table*}
\centering
\begin{tabular}{ccccccc}
\hline\hline
Time&Frequency& Band & \multicolumn{2}{c}{Avg PDS}  & \multicolumn{2}{c}{Avg LS periodogram} \\
\hline
& & & Single & Combined & Single & Combined \\
\hline
\multirow{4}{*}{MJD60765}&\multirow{2}{*}{$\sim 0.03$Hz} & S & $>$4.3$\sigma$ & \multirow{2}{*}{$>$5.9$\sigma$} & $>$4.3$\sigma$ & \multirow{2}{*}{$>$5.9$\sigma$} \\
                        &   & X & $>$4.3$\sigma$ &                        & $>$4.3$\sigma$ & \\
    &\multirow{2}{*}{$\sim 0.06$Hz} & S & $>$4.3$\sigma$ & \multirow{2}{*}{$>$5.9$\sigma$} & 2.0$\sigma$ & \multirow{2}{*}{$>$4.5$\sigma$} \\
                            &   & X & $>$4.3$\sigma$ &                        & $>$4.3$\sigma$ & \\
\hline
\multirow{4}{*}{MJD 60772}&\multirow{2}{*}{$\sim 0.03$Hz} & S & -- & \multirow{2}{*}{2.8$\sigma$} & 2.0$\sigma$ & \multirow{2}{*}{4.0$\sigma$} \\
                        &   & X & 2.8$\sigma$ &                        & 3.7$\sigma$ & \\
    &\multirow{2}{*}{$\sim 0.06$Hz} & S & 3.6$\sigma$ & \multirow{2}{*}{$>$5.4$\sigma$} & 3.5$\sigma$ & \multirow{2}{*}{5.0$\sigma$} \\
                        &   & X & $>$4.3$\sigma$ &                        & 3.8$\sigma$ & \\
\hline\hline
\end{tabular}
\caption{The significance of the $\sim$0.03 Hz QPO and its harmonic $\sim$0.06 Hz in averaged PDS and LS periodogram.}\label{tab:QPOsigma}
\end{table*}

\section{Discussion}

\subsection{Estimating Jet properties from radio flare}
\label{sec:jet}

The radio flare was observed simultaneously at X- (8.42~GHz) and S-band (2.25~GHz), with the X-band peak preceding the S-band peak. The spectral index $\alpha$ evolved from positive ($\alpha \sim +0.5$) during the flare rise to negative ($\alpha \sim -0.75$) after the peak, remaining roughly constant after MJD~60700. This behavior is characteristic of synchrotron self-absorption in an expanding jet \citep{Laan1966, Fender2019}: the emission is initially optically thick at low frequencies and becomes optically thin as the synchrotron emitting region expands and the optical depth decreases, causing the low-frequency peak to lag.

For the optically thin spectrum with $\alpha \simeq -0.75$, the electron energy distribution follows $N(E)\,{\rm d}E \propto E^{-p}{\rm d}E$ with $p \simeq 2.5$, typical of shock-accelerated synchrotron particles \citep{Spitkovsky2008ApJ}. The integrated 2 -- 8~GHz luminosity is $\sim1.5\times10^{33}\ {\rm erg\ s^{-1}}$. Assuming equipartition between particle and magnetic field energies \citep{Scott1977, Fender2019}, we estimate a minimum total jet energy of ($0.5$ -- $3.1) \times 10^{41}\ {\rm erg}$, a magnetic field strength of $(0.2$ -- $0.8)\ {\rm G}$, and a brightness temperature of $(5.5$ -- $5.8)\times10^{10}\ {\rm K}$, adopting a distance of 9.4~kpc \citep{Reid2023ApJ}. The inferred effective expansion velocity, assuming the rise began near MJD~60660, is $(1$ -- $3)\times10^{-3}c$ \citep{Fender2019}.

It is important to note that the derived minimum energy represents a conservative lower limit for this 2025 flare. Our weekly observational cadence, together with a substantial observational gap between MJD 60700 and MJD 60735, likely caused us to miss flares occurring on timescales of a few days or shorter, particularly if multiple short-duration flares were superposed on a longer one, as seen during the rapid radio activity of V404 Cyg \citep{Fender2023MNRAS}. In addition, a modest flux enhancement is visible around MJD 60700 (Figure~\ref{ref:flare}a), suggesting a subsequent radio flare whose rising phase was probably missed due to the data gap. If the detection on MJD 60699 approximately corresponds to the peak of this flare, we can crudely estimate its minimum energy to be $(0.8$--$6.5)\times10^{40} \mathrm{erg}$ following the method in \citet{Fender2019}. Consequently, the energy derived from our primary analysis accounts only for the initial radio flare and likely underestimates the total energy released during the 2025 observations.

\subsection{Possible origin of the 0.03Hz radio oscillation}
\label{sec:origin}

We detected a $\sim$0.03~Hz QPO in both X- and S-band observations on MJD~60765 and MJD~60772, respectively. To test for instrumental origins, we examined contemporaneous data from other sources. The calibrator scans were too short to produce PDSs, and no comparable peaks were found in the PDSs of 1ES~1927+654 observed on the same days. We further analyzed the \texttt{alist} files, which record fringe-fitting diagnostics sensitive to instrumental or atmospheric variations: Multi-Band Delay (clock-related), Single-Band Delay (ionospheric), and Delay Rate (frequency instabilities; \citealt{Thompson2017}). None of these showed significant power near 0.03~Hz. The absence of this signal in these diagnostics strongly indicates that the radio QPO is of astrophysical origin, most likely associated with the jet activity of GRS 1915+105 (see Section~\ref{sec:jet}).

\subsubsection{Accretion-driven scenarios: disk instabilities and accretion-ejection coupling}

X-ray QPOs across a wide frequency range have been observed in this source \citep[e.g.,][]{Morgan1997ApJ,Zhang2020MNRAS}. To explore possible correlations between radio and X-ray variability, we analyzed contemporaneous \textit{NICER} data. No \textit{NICER} observation exists on MJD~60765, and the two exposures on MJD~60772 ($>2000$~s each) showed only red-noise PDSs without any QPO near 0.03 or 0.06~Hz. Combining all \textit{NICER} data between MJD~60765 and 60772 likewise yielded no significant periodicity, likely due to strong X-ray obscuration suppressing variability detection \citep{Chen2025arXiv}.

Historically, GRS 1915+105 has exhibited recurrent radio and X-ray oscillations with periods of 20--40 min \citep{PF1997MNRAS, Mirabel1998, FenderP2000}, as well as much shorter X-ray oscillations \citep{Belloni2000}. Recently, a $\sim$8~min radio oscillation was detected during its current obscuration phase in 2023 \citep{RM2025ApJ}. Such behavior is generally attributed to disk-jet instabilities, in which cyclic evacuation and refilling of the inner disk lead to recurrent enhancements in accretion followed by ejection event \citep{Belloni1997ApJ, Mirabel1998, Vadawale2003}. Numerical simulations of such instabilities can reproduce simultaneous X-ray and radio oscillations on timescales from tens to thousands of seconds  \citep{Janiuk2000, Nayakshin2000, Neilsen2012}. An alternative explanation invokes the accretion-ejection instability (AEI) operating in the magnetized inner accretion disk \citep{Tagger1999,Tagger2004}. In this scenario, the gradual accumulation of poloidal magnetic flux in the inner disk, followed by rapid release once near equipartition, naturally leads to limit-cycle behavior characterized by repeated disk oscillations and ejections.

On the other hand, X-ray QPOs at frequency of tens of mHz have been reported in GRS~1915+105 \citep[e.g.][]{Morgan1997ApJ}. \citet{Malzac2018} demonstrates that accretion-flow variability observed in X-rays can propagate efficiently into the jet. However, the resulting QPO signal at longer wavelengths is expected to be weak, because internal shocks in the jet act as a filter that damps rapid variability. Therefore, in the absence of simultaneous X-ray timing observations, we cannot determine whether the detected radio QPO is directly linked to inner accretion-flow variability, regardless of the underlying physical mechanism.

\subsubsection{Jet-related scenarios: intrinsic jet instabilities and jet precession}
It is also plausible that the observed radio QPO originates intrinsically within the jet, without requiring a periodic driver from the accretion flow.  
For instance, three-dimensional relativistic MHD simulations demonstrate that kink instabilities can trigger the quasi-periodic dissipation of magnetic energy in the jet spine, generating QPO signatures in both flux and polarization \citep{Dong2020}. Notably, this model successfully accounts for the observed anti-correlation between flux and polarization degree observed during the $\sim$0.03 Hz QPO episode reported by \citet{Wang2025NC}. However, the kink instability is inherently transient and stochastic, with a period determined by the local instability growth timescale. 
As such, the recurrence of the $\sim$0.03~Hz frequency across multiple epochs in 2021 \citep{Wang2025NC} and 2025 (\autoref{ref:flare}) is inconsistent with a stochastic origin, making it improbable that kink instability alone sustains this oscillation at a constant frequency.

An alternative and appealing interpretation of the detected radio QPO is Lense-Thirring (LT) precession of the jet \citep{LT1918PhyZ,Stella1998,Malzac2018,Ma2021}. Such rapid jet precession has been evidenced in the BHXRB V404~Cygni, where minute-timescale changes in jet orientation were directly imaged \citep{Miller2019}. General relativistic magnetohydrodynamic (GRMHD) simulations also demonstrate that the jet is magnetically coupled to the inner accretion flow and can be forced to precess coherently with it \citep[e.g. ][]{Fragile2007,McKinney2013,Liska2018,Liska2019,Liska2021,Chatterjee2025}.
 
In addition, quasi-periodic modulations not only in flux, but also in the linear polarization degree and polarization angle are expected if the jet is dynamically coupled to and co-precesses with the inner accretion flow \citep{Ingram2015}. The $\sim$0.03 Hz QPO has been detected simultaneously in both radio flux and linear polarization during the 2021 flare of GRS~1915+105 \citep{Wang2025NC}, at a frequency consistent with our result. However, the oscillation in polarization angle is detected at $\sim$0.025~Hz with only marginal statistical significance \citep{Wang2025NC}.

Moreover, GRS 1915+105 has remained in a prolonged X-ray obscured phase since 2019, likely due to a warped disk that blocks the central X-ray region \citep[e.g. ][]{Miller2025}. Such a disk geometry tends to sustain a misalignment between the angular momentum of accretion disk and the black hole, creating conditions favorable for LT precession.
In this context, the observed radio QPO is expected to arise from LT-driven precession of a jet that is magnetically coupled to the inner accretion flow.

If the observed QPO indeed originates from the LT precession of the jet, for a BH mass of $M_\mathrm{BH}=11.2\,M_{\odot}$ \citep{Reid2023ApJ} in GRS 1915+105, the 0.03~Hz frequency corresponds to a characteristic radius of $\sim$55~$R_\mathrm{g}$ for a spin value of $a_* \simeq 0.98$ \citep{Reid2014ApJ} and $\sim$50~$R_\mathrm{g}$ for a spin value of $a_* \simeq 0.7$ \citep{Motta2023}. This radius is consistent with the disk-jet co-precessing region predicted in the aforementioned GRMHD simulations \citep[e.g. ][]{McKinney2013,Liska2018,Liska2021}. 
Evidence for such a co-precession region has recently emerged from the synchronized X-ray and radio quasi-periodic variability ($\sim$19.6 days) observed in the tidal disruption event AT2020afhd \citep{Wang2025NaturTDE}, and similar ideas have been invoked to explain the quasi-periodic position angle variations ($\sim$11.24 years) of the radio jet in the supermassive black hole M87 \citep{Cui2023,Cui2025}. Together, these results may point to a common physical mechanism operating across a wide range of black hole masses and accretion regimes, although the characteristic size of the inferred co-precession region differs among them.

\section{Summary}

We conducted a weekly-cadence radio monitoring campaign for the microquasar GRS 1915+105, carried out simultaneously at 2.25 GHz (S band) and 8.42 GHz (X band). 
Our observations captured a bright radio flare characterized by a clear spectral evolution from optically thick to optically thin synchrotron emission, consistent with an expanding relativistic jet. From high time resolution analysis (2-second bins), we detected a significant QPO at $\sim$0.03~Hz and its harmonic at 0.06~Hz in two consecutive observations on MJD~60765 and 60772. The joint significance calculated by combining S and X bands exceeds $5\sigma$.

A critical finding is the remarkable recurrence and wavelength independence of the QPO frequency. The 0.03 Hz oscillation was identical in both the S- and X-bands. Although these wavelengths probe different characteristic emission regions along the jet, the wavelength-independent oscillation suggests a global modulation mechanism operating coherently across multiple jet scales. Furthermore, this characteristic frequency matches a 0.03 Hz QPO previously detected in a separate radio flare event from January 2021 \citep{Wang2025NC}, demonstrating that this oscillation frequency has remained constant over a multi-year baseline across different flare episodes. The recurrence of this QPO frequency suggests it reflects an intrinsic characteristic timescale of the accretion-jet system. Yet, its detection in only two consecutive observations over 100 days of flare episode implies the excitation mechanism is transient, such as specific jet geometry, magnetization, or accretion state. 
Additionally, while several mechanisms could plausibly explain this oscillation, the heavy X-ray obscuration currently precludes a more in-depth exploration. To map these physical conditions and ultimately confirm the origin of the QPO, future simultaneous X-ray timing and high spatial resolution radio imaging on short timescales will be essential to track jet structural evolution during such episodes.


\begin{acknowledgments}
The authors thank the anonymous referee for their constructive comments and suggestions that helped improve this paper. We also thank the telescope operators and scheduling staff at SHAO for their essential work. Y.Z. and L.Y.X. thank Wenfei Yu and Agnieszka Janiuk for helpful discussions. This work is partially supported by Natural Science Foundation of China (NSFC) grants 12373049, 12361131579, 12173074). L.L. is partially supported by National Key Research and Development Program of China (grant No. 2022YFC2205203).
\end{acknowledgments}




%
\facilities{Tianma radio telescope, Sheshan radio telescope, NICER}

\software{DiFX \citep{Deller2007PASP},  
          Stingray \citep{Huppenkothen2019ApJ}
          }





\bibliography{ref}{}
\bibliographystyle{aasjournalv7}


\end{CJK*}
\end{document}